\documentclass[prd,amsmath,amsfonts,preprint,12pt,superscriptaddress]{revtex4}
\usepackage{epsfig}

\begin{document}

\title{Kasner and Mixmaster behavior in universes with \\
equation of state $w\ge1$}
\author{Joel K. Erickson}
\affiliation{Joseph Henry Laboratories, Princeton University, Princeton NJ, 08544}
\author{Daniel H. Wesley}
\affiliation{Joseph Henry Laboratories, Princeton University, Princeton NJ, 08544}
\author{Paul J. Steinhardt}
\email{steinh@princeton.edu}
\affiliation{Joseph Henry Laboratories, Princeton University, Princeton NJ, 08544}
\affiliation{School of Natural Sciences, Institute for Advanced Study,
Olden Lane, Princeton, NJ, 08540}
\author{Neil Turok}
\affiliation{DAMTP, CMS, Wilberforce Road, Cambridge, CB3\,0WA, UK}

\begin{abstract}
We consider cosmological models with a scalar field with equation of
state $w\ge 1$ that contract towards a big crunch singularity, as in
recent cyclic and ekpyrotic scenarios. We show that chaotic mixmaster
oscillations due to anisotropy and curvature are suppressed, and the
contraction is described by a homogeneous and isotropic Friedmann
equation if $w>1$.  We generalize the results to theories where the
scalar field couples to $p$--forms and show that there exists a finite
value of $w$, depending on the $p$--forms, such that chaotic
oscillations are suppressed. We show that $\mathbb{Z}_2$ orbifold
compactification also contributes to suppressing chaotic behavior. In
particular, chaos is avoided in contracting heterotic $M$-theory
models if $w>1$ at the crunch.
\end{abstract}

\maketitle


\section{Introduction}

In cosmological models with a big crunch/big bang transition, a key
issue is the behavior as the universe contracts towards the crunch.
From the classic studies of Belinskii, Khalatnikov and Lifshitz (BKL)
\cite{Lif63, Bel70,Bel73} and others
\cite{Dem85,Dam00,Dam00A,Dam02,Dam02A}, it is known that the
contraction can either proceed smoothly or chaotically.  These studies
have focused on models in which the universe contains matter and
radiation, or, more generally, an energy component whose equation of
state is $w\le 1$ (where $w\equiv p/\rho$ is defined as the ratio of
the pressure $p$ to the energy density $\rho$).  If $w<1$, a
contracting homogeneous and isotropic solution is unstable to small
perturbations in the anisotropy and spatial curvature.  As the overall
volume shrinks, the anisotropy causes the universe to expand along one
axis and contract along the others, a state that can be approximated
by the anisotropic Kasner solution.  The spatial curvature causes the
axes and rates of contraction to undergo sudden jumps from one
Kasner-like solution to another, an effect known as ``mixmaster''
\cite{Mis69,MTW} behavior.  If the curvature is not spatially uniform,
then the chaotic behavior in different regions is not synchronized and
the universe becomes highly inhomogeneous at the big crunch.  Hence,
mixmaster behavior could potentially wreak havoc in cosmological
models with a big crunch/big bang transition, making them inconsistent
with the observed large scale homogeneity of the universe.

In this paper, we show that the behavior of the universe as it approaches the big 
crunch is very different if there is an energy component with $w>1$.  The 
chaotic behavior is suppressed and the universe contracts homogeneously and 
isotropically as it approaches the singularity.  The reason is that the 
anisotropy and curvature terms in the Einstein equations grow rapidly and 
become dominant if $w<1$, but they remain negligible compared to the energy 
density if $w>1$.  
In the latter case, the Einstein equations converge to the 
Friedmann equations with purely time-dependent terms, a condition sometimes 
referred to as ``ultralocality.'' The effect can be viewed as a generalization of 
the ``cosmic no-hair theorem'' invoked in a rapidly inflating universe. Here
we
demonstrate  analogous behavior in 
a slowly contracting universe with $w>1$.  
A related result of Dunsby et al \cite{Dun03,Col01A,Col01B} shows
that models with $0<w<1$ but with $\rho^2$ terms in the stress-energy tensor
are also driven towards isotropy.

The cosmic no hair theorem for a contracting 
universe containing a perfect fluid with $w \ge 1$ is discussed in section \ref{sec:nohair}.  A 
common example of a perfect fluid is a scalar field $\phi$ with a  
potential $V(\phi)$.  In section \ref{sec:pforms}, we consider the interaction of the 
scalar $\phi$ with 
a $p$--form field $F_{p+1}$ through an exponential coupling,
\begin{equation}
e^{\lambda \phi} \, F_{p+1}^2,
\end{equation}
where $\lambda$ is a constant \cite{note}. We consider this case because
scalar fields with exponential couplings to $p$--form fields are common in Kaluza-Klein, supergravity and superstring models.  For the case $w=1$, it is
known \cite{Bel73,Dam02} that  the contraction is not chaotic if $\lambda$ lies within a bounded interval.  
Here we show that, for {\it any } $\lambda$ and $p$, there is a
critical value $w_{\text{crit}}(\lambda,p)$ for which the chaotic behavior is
suppressed if $w> w_{\text{crit}}(\lambda,p)$.

Our results are of particular importance for the recent ekpyrotic
\cite{Kho01A} and cyclic \cite{Ste02} cosmological models, which have
a big crunch/big bang transition with a contraction phase dominated by
a scalar field with $w\ge 1$ \cite{design}. 
The evolution of perturbations leading up to and passing through the
transition is an important aspect that remains unsettled
\cite{Kal01A,Kal01B,Bra01,Lyt01A,Lyt01B,Mar01,Mar02}, and  may depend
on the precise physical conditions leading up to the bounce
\cite{Tol03,Bra04}. The present work may be relevant since it suggests that
the universe can remain homogeneous and isotropic on large scales.
Once the evolution becomes ultralocal, the whole universe
is following the same homogeneous and isotropic evolution all the way
to the big crunch.

In section \ref{sec:time_vary}, we explore how time-variation of $w$
affects our conclusions, and in particular how $w$ approaching
$w_{\text{crit}}$ from above may suppress chaotic behavior. In section
\ref{sec:extraD} we discuss some specific models. In particular, we
show how orbifolding can remove $p$-forms that might induce chaotic
behavior and discuss the special case of heterotic $M$-theory, which,
to leading order in the eleven dimensional gravitational coupling
$\kappa$, is on boundary between chaotic and smooth behavior.

\section{A ``Cosmic No--Hair Theorem'' for Contracting Universes}\label{s:no_hair}
\label{sec:nohair}

The cornerstone of the inflationary paradigm is an argument known as
the ``cosmic no-hair theorem'', according to which a universe
containing a perfect fluid component with $w < -1/3$ will rapidly
approach flatness, homogeneity and isotropy at late times, for a wide
range of initial data (namely those for which the space curvature,
inhomogeneity and anisotropy are not very large) \cite{KT90}.  In the
Friedmann equation, the energy density for a component with equation
of state $w$ is proportional to $1/a^x$, where the exponent
$x=3(1+w)$. The anisotropy term is proportional to $a^{-6}$ and the
spatial curvature term is proportional to $a^{-2}$. As the universe
expands, the contribution with the smallest values of $x$ redshifts
away more slowly than components with larger values of $x$ and so come
to dominate the Friedmann equation and the components with the
smallest value of $x$ overall ultimately dominate. If the energy
component with the smallest value of $w$ has $w<-1/3$, then $x<2$ and
this component dominates. For a wide range of initial data,
convergence to a homogeneous and isotropic expanding universe is
assured.

Below, we will present an analogous ``cosmic no-hair theorem'' for
contracting universes. In a contracting universe, the component with
the largest value of $x$ will dominate the Friedmann
equation. Starting from an inhomogeneous and anisotropic initial
state, we will show that the existence of a perfect fluid with $w > 1$
(or $x>6$) will suppress chaotic behavior, and enable a smooth and
isotropic contraction to the big crunch.  We will find that curvature
plays a more complicated role compared to the case of
expansion. Hence, we first obtain a cosmic no-hair theorem for the
case of zero spatial curvature and then generalize to the case of
arbitrary spatial curvature. We intentionally take a pedagogical
approach that encompasses known results for $w \le 1$ to make our
discussion self-contained. 
Our analysis assumes the initial inhomogeneity is small; it is possible that the  universe evolves
towards other attractors for sufficiently large deviations from
homogeneity.
Our conventions are given in \cite{note}.

All of our computations are performed in synchronous gauge,
\begin{equation}
ds^2=-dt^2+h_{ab}(t,\mathbf{x})\,dx^a\,dx^b,
\label{eq:synchronous}
\end{equation}
where we use our freedom to choose a spatial slicing to ensure that
the big crunch occurs everywhere at $t=0$ ($\det h_{ab}\to0$ as
$t\to0$). For a perfect, comoving fluid with equation of state $p =
w\rho$, the Einstein equations are \cite{Lif63}:
\begin{subequations}
\label{eq:einstein}
\begin{align}
\frac{\partial}{\partial t} {\kappa_j}^j + 
{\kappa_j}^k{\kappa_k}^j & = 
-\,\Bigl( \frac{1+3w}{2} \Bigr) \rho, \label{eq:R00} \\
\frac{\partial}{\partial x^a}{\kappa^j}_j-
\frac{\partial}{\partial x^j}{\kappa^j}_a & = 0, \label{eq:R0a}\\
{P_a}^b + \frac{1}{\sqrt{h}}\frac{\partial}{\partial 
t}\Bigl(\sqrt{h}{\kappa_a}^b\Bigr) & = 
+\Bigl( \frac{1-w}{2} \Bigr) \rho \label{eq:Rab},
\end{align}
\end{subequations}
where ${P_a}^b$ is the Ricci tensor on spacelike
surfaces, and $\kappa_{ab}$ is defined by
\begin{subequations}
\begin{align}
\kappa_{ab}&=\frac{1}{2}\frac{\partial}{\partial t}h_{ab},\\
{\kappa_a}^b&=\kappa_{aj}h^{jb}.
\end{align}
\end{subequations}

Near the big crunch, the dynamics of the metric (\ref{eq:synchronous})
are \textit{ultralocal} \cite{Bel70,Dam02,Dam02A,And00}. That is, the
evolution of adjacent spatial points decouples because spatial
gradients increase more slowly than other terms in the equations of motion.
Therefore, analyzing the
dynamics of this metric near the singularity and at fixed spatial
coordinate $\mathbf{x}_0$ is equivalent to analyzing the much simpler
system
\begin{equation}
ds^2=-dt^2+\sum_{ij}e^{2\beta_{ij}(t;\mathbf{x}_0)}\sigma^{(i)}(\mathbf{y};\mathbf{x_0})\sigma^{(j)}(\mathbf{y};\mathbf{x}_0),
\end{equation}
where the $\sigma^{(i)}$ are $\mathbf{y}$-dependent one-forms that are
linearly independent at each point and form a \textit{homogeneous}
(but possibly curved) space such as Bianchi type IX \cite{MTW}. The
$\beta_{ij}$, which do not depend on $\mathbf{y}$, describe the
(generally anisotropic) contraction of this space. Both the
$\sigma^{(i)}$ and the $\beta_{ij}$ depend on the parameter
$\mathbf{x}_0$, the spatial point being studied. The dynamics of the
inhomogeneous universe at a fixed spatial point can be approximated,
near $t=0$, by the dynamics of a homogeneous (but curved and
anisotropic) universe. Differences in curvature and anisotropy between
different $\mathbf{x}_0$ are encoded in the different $\sigma^{(i)}$
and $\beta_{ij}$ associated with these points.

In each Kasner-like epoch, we may perform a rotation so that
$\beta$ is diagonal. Furthermore, we may separate out the trace
of $\beta$ and write it as the ``volume scale-factor'' $a(t)$, in
analogy to the isotropic Friedman-Robertson-Walker universe, to obtain
the metric
\begin{subequations}
\begin{gather}
ds^2=-dt^2+a^2(t)\sum_i e^{2\beta_i(t)}(\sigma^{(i)})^2,\\
\beta_1(t)+\beta_2(t)+\beta_3(t)=0,\label{eq:beta_constraint}
\end{gather}
\label{eq:gen_kasner}
\end{subequations}
where the dependence of $a(t)$, the $\beta_i$ and the $\sigma^{(i)}$
on $\mathbf{x}_0$ has been suppressed.  The combination
$a\,e^{\beta_i}$ can be thought of as the effective scale factor along
the $i^{\text{th}}$ direction, and the functions $\beta_i$ then
describe the contraction or expansion of each direction relative to
the overall volume contraction.  We may use our freedom to rescale the
$\sigma$ to ensure that at some time $t_0$, $a(t_0)=1$,
$\beta_i(t_0)=0$ and
$\det(\sigma^{(1)},\sigma^{(2)},\sigma^{(3)})=1$. Quantities with a
subscript zero (such as $\rho_0$) refer to their values at this fixed
time.

The Einstein equations (\ref{eq:einstein}) close with the equation
of energy conservation for the fluid,
\begin{equation}
\frac{d \log \rho}{d \log a} = -3 (1+w).
\label{eq:conservation}
\end{equation}
For constant $w$, this equation has the familiar solution,
\begin{equation}
\rho(a) = \rho_0 a^{-3(1+w)}.
\label{eq:rho_a}
\end{equation}
While we could have included several perfect fluids, with different
equations of state $w_i$, the fluid with the largest equation of state
will always dominate near the crunch, so it is sufficient to consider
only one energy component. We have taken this fluid to be comoving,
because small perturbations of a comoving background are suppressed in
a $w>0$, contracting universe. In particular, the $T^0{}_i$ terms that
would appear on the right hand side of (\ref{eq:R0a}) grow only as
$t^{-2/(1+w)}$, which is slower than the $t^{-2}$ rate at which the
diagonal terms grow \cite{noteA}.

\subsection{The Curvature--Free Case}
\label{sec:nocurv}

We first examine the case of Ricci flat spatial $3$--surfaces, for which
$ {P_a}^b = 0$. In this case, we write $\sigma^{(i)}=dx^i$.
Then, the Einstein equations (\ref{eq:einstein}) reduce to
\begin{subequations}
\begin{align}
3\Bigl(\frac{\dot{a}}{a}\Bigr)^2-
\frac{1}{2}(\dot\beta_1^2+\dot\beta_2^2
+\dot\beta_3^2)&= \rho, \label{eq:mod_einstein_1}\\
\ddot\beta_i+3\frac{\dot{a}}{a}\dot\beta_i &=0 \label{eq:mod_einstein_2},
\end{align}
\end{subequations}
where a dot indicates a derivative with respect to the proper time $t$.
Integration of (\ref{eq:mod_einstein_2}) gives,
\begin{equation}
\dot \beta_i = c_i a^{-3},
\end{equation}
while the constraint (\ref{eq:beta_constraint}) implies,
\begin{equation}
c_1 + c_2 + c_3 = 0.
\label{eq:kasner_us_1}
\end{equation}
Combining these results, equation (\ref{eq:mod_einstein_1}) becomes a
Friedmann equation,
\begin{equation}
3\Bigl(\frac{\dot a}{a}\Bigr)^2 =  \rho(a) +
 \frac{\sigma^2}{a^{6}} = \frac{\rho_0}{a^{3(1+w)}} +
\frac{\sigma^2}{a^{6}},
\label{eq:mod_fried}
\end{equation}
where we define
\begin{equation}
\sigma^2 = \frac{1}{2}( c_1^2 + c_2^2 + c_3^2).
\label{eq:kasner_us_2}
\end{equation}
An anisotropic universe has $\dot{\beta}_i \ne 0$, \textit{i.e.} $c_i \ne 0$. 
The constant $\sigma^2$ parameterizes the anisotropic contribution to
the Friedmann equation in (\ref{eq:mod_fried}). The
anisotropy evolves as $1/a^6$ or $x=6$. We define the fractional
energy densities $\Omega_\rho$ and $\Omega_\sigma$ as
\begin{subequations}
\begin{align}
\Omega_\rho&=\frac{\rho(a)}{\rho(a)+\sigma^2/a^6},\\
\Omega_\sigma&=\frac{\sigma^2/a^6}{\rho(a)+\sigma^2/a^6}.
\end{align}
\end{subequations}
These quantities represent the contribution of the perfect fluid and
anisotropy to the critical density for closure of the universe. Since
we are neglecting curvature, $\Omega_\rho+\Omega_\sigma=1$. 

The solution for the $\beta_i$ as a function of the 
scale factor $a$ is,
\begin{equation}
\beta_i(a) = c_i \sqrt{3} \int_a^1\frac{da'}{a'}\bigl( \rho(a')
a'^6 + \sigma^2\bigr)^{-1/2}.
\label{eq:beta_w_of_a}
\end{equation}
The limits of integration have been chosen to ensure
$\beta_i(1)=0$. For the remainder of the paper, we will assume a
universe contracting towards $a \to 0$ as $t$ approaches zero from
below.  Let us now examine the behavior of these solutions for various
$w$.

\subsubsection{$w < 1$:}

\begin{figure}
\epsfig{file=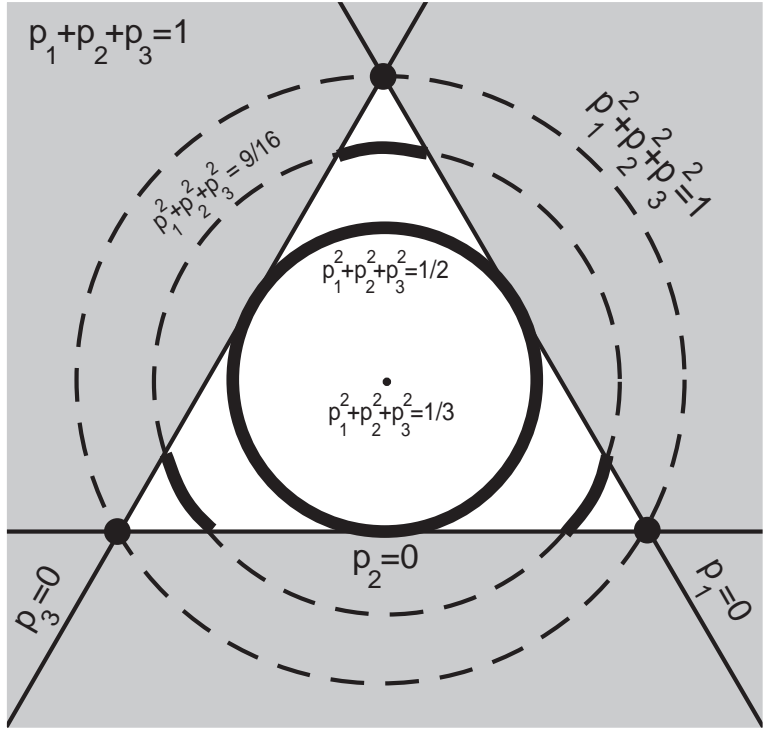,width=4in}
\caption{The Kasner plane $p_1+p_2+p_3=1$ and its intersections (the 
Kasner circles) with various 
spheres $p_1^2 + p_2^2 + p_3^2 = 1 -q^2$
where $q^2=\tfrac{2}{3}(1-\Omega_\sigma)$; see (\ref{eq:q2}).  
The vacuum solution corresponds to $\Omega_\sigma=1$ (the outermost circle).
The inner circles are relevant to the case where 
$w=1$ and $\Omega_\sigma<1$.
In the white regions, the Kasner exponents are all positive
(corresponding to contraction); in gray regions, one exponent is 
negative (expanding).  If the spatial curvature is non-zero, 
 points along  the circles in the white region 
(thick parts of circles) are stable but
 points in the gray regions (dashed parts of circles) are unstable, 
jumping to new values after a short period of contraction.
If a model (\textit{i.e.}~a circle)   has an open set of 
stable points  (the three innermost circles but not the
outermost circle), the contracting
phase does not exhibit chaotic mixmaster behavior. }
\label{fig:kasner}
\end{figure}

When $w < 1$, the $\rho(a)$ part of the
integral (\ref{eq:beta_w_of_a})
is negligible as $a\to 0$, and so the solution
converges to the vacuum ($\rho = 0$) Kasner universe during contraction,
\begin{subequations}
\begin{gather}
a(t) = \Bigl(\frac{t}{t_0}\Bigr)^{1/3} \\
\beta_i(t) = \frac{c_j}{\sigma\sqrt{3}} \ln{ \Bigl(\frac{t}{t_0}\Bigr)}.
\end{gather}
\label{eq:soln_rho_0}
\end{subequations}
The Kasner universe is parameterized by three \emph{Kasner exponents}
$p_i$,
\begin{equation}
p_i = \frac{1}{3} + \frac{c_j}{\sigma\sqrt{3}}.
\label{eq:us_2_kasner}
\end{equation}
The scale factors in (\ref{eq:gen_kasner}) are powers
of $t$:
\begin{equation}
ae^{\beta_i} = |t/t_0|^{p_j},
\end{equation}
and the relations (\ref{eq:kasner_us_1}) and (\ref{eq:kasner_us_2}) become
\begin{subequations}
\begin{align}
p_1 + p_2 + p_3 &= 1 \label{eq:kasner_1}\\
p_1^2 + p_2^2 + p_3^2 &= 1 \label{eq:kasner_2},
\end{align}
\end{subequations}
known as the \textit{Kasner conditions}.  These describe the
intersection of a plane, the \textit{Kasner plane}, and a unit sphere,
the \emph{Kasner sphere}, as illustrated in Fig.~\ref{fig:kasner}.  We
will denote the intersection, which represents the allowed values of
the $p_i$, as the \textit{Kasner circle}.  The outermost circle in
Fig.~\ref{fig:kasner} corresponds to the limit where $w<1$, as
the energy density scales away and only a vacuum, anisotropic universe
remains.

There are three degenerate solutions where exactly one of the $p_i$ is
one, and the other exponents are zero (the solid black circles in
Fig.~\ref{fig:kasner}).  At all other points on the (dashed) Kasner
circle exactly one of the $p_i$ is negative.  Thus, although the
geometric mean of the three scale factors $a(t) = |t|$ is contracting,
a single scale factor corresponding to the negative Kasner exponent is
undergoing expansion to infinity.

For the curvature-free case, the universe becomes increasingly
anisotropic near the big crunch if $w<1$.  In particular, the
isotropic solution, $p_1= p_2=p_3=1/3$, is inconsistent with the
Kasner conditions (\ref{eq:kasner_1}) and (\ref{eq:kasner_2}).

\subsubsection{$w = 1$:}
Inspection of (\ref{eq:rho_a}) reveals that, when $w = 1$, the matter
density and the anisotropy terms in the Friedmann equation
(\ref{eq:mod_fried}) scale with the same power of $a$, so
$\Omega_\rho$ and $\Omega_\sigma$ remain fixed.  The solutions are
\begin{subequations}
\begin{gather}
a(t) = \Bigl(\frac{t}{t_0}\Bigr)^{1/3}\\
\beta_i(t) = \frac{c_j }{\sqrt{3(\sigma^2 + \rho_0)}} \ln{ \Bigl(\frac{t}{t_0}\Bigr)}.
\end{gather}
\end{subequations}
This solution is very similar to the $\rho = 0$ case, and indeed we
may define the Kasner exponents,
\begin{equation}
p_i = \frac{1}{3} + \frac{c_j}{\sigma\sqrt{3}} \Bigl( 1 + \frac{\rho_0}{\sigma^2} \Bigr)^{-1/2}.
\label{eq:us_2_kasner_2}
\end{equation}
The Kasner conditions are different.  If we define
\begin{equation}
q^2 \equiv \frac{2}{3} \frac{\rho_0}{\sigma^2 + \rho_0}=\frac{2}{3}(1-\Omega_\sigma)
\label{eq:q2}
\end{equation}
then the Kasner conditions are
\begin{subequations}
\begin{align}
p_1 + p_2 + p_3 &= 1, \\
p_1^2 + p_2^2 + p_3^2 &= 1 - q^2=\tfrac{1}{3}+\tfrac{2}{3}\Omega_\sigma.
\end{align}
\end{subequations}
The first condition is unchanged from (\ref{eq:kasner_2}) but the
right hand side of the second condition has been modified.  Increasing
$\Omega_\sigma$ corresponds to increasing the radius of the Kasner sphere.

The $w=1$ model allows us to explore the behavior of the contracting
universe as a function of $\Omega_\sigma$.
The perfectly isotropic case corresponds
to $\Omega_\sigma=0$, which is the usual flat
Friedmann-Robertson-Walker solution (innermost circle, in the limit
where the circle has shrunk to a point, in Fig.~\ref{fig:kasner}).
Unlike the vacuum Kasner case, all of the Kasner exponents are
positive (\textit{i.e.}~lie within the white region of Fig.~\ref{fig:kasner})
provided that $\Omega_\sigma<1/4$ (within the larger, solid circle
inscribed in the triangle).  For this range, none of the scale factors
is increasing during the contraction, although they are decreasing at
different rates.  When $\Omega_\sigma>1/4$ (third largest circle),
then some points on the Kasner circle have a negative Kasner exponent
(dashed part of circle) and other points may have all positive Kasner
exponents (solid, thick parts of circle).

Thus, ignoring the curvature, the $w = 1$ case with non-zero
$\Omega_\sigma$ contracts smoothly but anisotropically to the crunch.
In the special case where $\Omega_\sigma=0$, the contraction is
isotropic.

\subsubsection{$w > 1$:}
For $w>1$, the energy density dominates ($\Omega_\rho\to 1$) as
$a\to 0$, and the metric approaches the approximate form
\begin{subequations}
\begin{gather}
a(t) = \Bigl(\frac{t}{t_0}\Bigr)^{2/3(1+w)}, \\
\beta_i(t) = c_j\frac{2}{\sqrt{3\rho_0}} \frac{1}{w-1}
\Bigl[\Bigl( \frac{t}{t_0} \Bigr)^{\frac{w-1}{w+1}}-1\Bigr],
\end{gather}
\label{eq:wg1}
\end{subequations}
where we have chosen the constants of integration so $\beta_i=0$ at
$t=t_0$.  The crucial feature is that the time-varying part of the
$\beta_i$ is proportional to $t^{\alpha}$ where $\alpha$ is {\it
positive} if $w>1$. This means that the $\beta_i$ approach a constant
and the universe becomes isotropic at the crunch \cite{noteB}.

\medskip

This simple result is a ``no--hair theorem'' for universes without
spatial curvature: When $w > 1$, an initially anisotropic universe
becomes isotropic ($\Omega_\sigma\to0$) near the big crunch. The $w>1$
case is stable under anisotropic perturbations.  For $w < 1$, the
universe becomes increasingly anisotropic in the sense that
$\Omega_\sigma\to 1$ as $a\to 0$.  For $w=1$, $\Omega_\sigma$ remains fixed
as $a \to 0$.  Evolution is smooth (no mixmaster behavior) in all
cases, and is well-approximated as a Kasner metric with constant coefficients
for sufficiently small $a$.

\subsection{Curvature and Chaos}
\label{sec:curv}

Complex behavior can arise when there is non-zero spatial curvature in
a contracting universe.  This may seem surprising at first, since the
spatial curvature for a homogeneous and isotropic universe grows as
$1/a^2$, which increases more slowly than
either the anisotropy or the energy density of a
component with $w
>-1/3$.  However, we have seen above that the contracting
phase for $w\le 1$ is anisotropic.  We will show below that this
can produce rapidly growing curvature perturbations and chaotic
behavior. On the other hand, we will see that chaotic behavior is
suppressed if $w>1$ and the contraction approaches isotropy as $a
\to 0$.

We now allow the $\sigma^{(i)}$ to have an $\mathbf{x}$-dependence
and consider a curved manifold. The spatial Ricci tensor for the
metric (\ref{eq:gen_kasner}) has the form \cite{Lif63}
\begin{equation}
{P_a}^b = \frac{1}{a^2} \sum_{ijk} {{S_a}^b}_{ijk}(\sigma) 
e^{  2( \beta_i - \beta_j - \beta_k) }.
\label{eq:pab}
\end{equation}
The functions ${{S_a}^b}_{ijk}$ depend only on the $\sigma^i$ and
their space derivatives, and are independent of time.

The expression (\ref{eq:pab}) reveals a crucial connection between the
behavior of anisotropy and curvature near the big crunch.  In the
isotropic limit, $\beta_i = 0$ and (\ref{eq:pab}) reduces to the
homogeneous and isotropic $1/a^2$ scaling discussed above.  However,
the terms in (\ref{eq:pab}) are essentially ratios of scale factors.
Thus, if the anisotropy is growing as $a \to 0$, some terms --
involving ratios of expanding and contracting scale factors -- will
grow, and the corresponding curvature components will scale faster
than $1/a^2$.  For $w < 1$ the anisotropy dominates near the crunch,
and, as we will discuss below, this causes the curvature to grow and
induce chaos.  By contrast, in the $ w > 1$ model, the anisotropy
vanishes at the crunch, and the curvature scales as the usual $1/a^2$,
which may be neglected.

\subsubsection{$w < 1$:}

In this case, we begin by assuming that the behavior near the crunch 
is described by the
vacuum Kasner solution, with Kasner conditions (\ref{eq:kasner_1}) and
(\ref{eq:kasner_2}).  Using the Kasner solution, it is readily seen that the
Einstein equation (\ref{eq:R00}) contains a leading order term with time
dependence $t^{-2}$.  

The second Einstein equation (\ref{eq:R0a}) is a consistency check for
our assumption of ultralocality. For an appropriate choice of the
$\sigma^{(i)}$ -- a basis for one of the Bianchi universes -- this
equation vanishes identically and the metric (\ref{eq:gen_kasner})
solves the Einstein equations.

The third Einstein equation (\ref{eq:Rab}) indicates that the simple
Kasner solutions must break down near the big crunch.  If we order the
Kasner exponents as $p_1 \le p_2 \le p_3$, then the most divergent
term in the third Einstein equation comes from terms in the spatial
curvature (\ref{eq:pab}), with leading time dependence,
\begin{equation}
t^{-2(1-2p_1)}.
\end{equation}
The leading term is more divergent than $t^{-2}$, since the Kasner
conditions (\ref{eq:kasner_1}) and (\ref{eq:kasner_2}) imply $p_1$ is
always negative.  Therefore, our smoothly contracting solutions are
not stable to perturbations in the spatial $3$--curvature.
A small amount of curvature will grow and come to dominate the
dynamics before the big crunch.

The behavior of the universe in this regime has been extensively studied
and is known to be  
chaotic \cite{Bel70,Bel73,Dem85,Dam00,Dam00A,Dam02A}.
The spatial curvature terms cause the Kasner exponents $p_i$ and the principal 
directions $\sigma^i$ to become time-dependent during contraction.  

More precisely, the exponents and principal directions are nearly
constant for stretches of Kasner-like contraction, during which the
curvature is negligible.  These Kasner-like epochs are punctuated by
short intervals when the curvature momentarily dominates. The
exponents and principal directions suddenly jump to new values, and
then a new stretch of Kasner-like contraction begins during which the
curvature terms are again negligible. The universe undergoes an
infinite number of such jumps before the big crunch. The chaotic,
non-integrable evolution is equivalent to that of a billiard ball
\cite{Dam02A}, which experiences free motion interrupted by collisions
with walls. Models with this oscillatory behavior are called
\textit{chaotic}.

This presents a problem for cosmological models, as one expects
curvature perturbations in any realistic universe will cause the local
value of the curvature to vary from point to point.  If each spatial
point evolves independently and chaotically, the evolution of nearby
points diverges very quickly as contraction continues, and the
universe rapidly becomes highly inhomogeneous as $a \to 0$. If $w<1$
throughout the contracting phase, it seems unlikely that the observed
homogeneous universe could emerge from this state after the bounce to
an expanding phase \cite{JP72}.

\subsubsection{$w = 1$:}

The chaotic behavior is mitigated in the $w = 1$ case.  Recalling our
discussion of the curvature-free scenario, it is clear that there are
regions of non-zero measure on the Kasner circle for which all of the
$p_i$ are positive. We will refer to these points as \textit{stable}.
All choices of $p_i$ when $\Omega_\sigma < 1/4$ are stable.  If
the universe begins at a stable point, the curvature term remains
negligible as $a \to 0$ and the contraction is smoothly
Kasner-like.

However, when $\Omega_\sigma > 1/4$, some choices of the $p_i$
will have one $p_i < 0$.  If the universe begins at one of these
points, the curvature term will grow and become dominant, causing the
values of $p_i$ and the principal axes $\sigma_i$ to change. We
refer to these points as \textit{unstable}. A more complete
analysis \cite{Bel73} reveals that, after a finite number of jumps,
the universe hits a point in the open set of stable $p_i$.  From this
point onwards, the universe contracts smoothly and without any further jumps.

We call these models \textit{non-chaotic}, since the universe is
guaranteed to arrive at a stable point as $a\to 0$. Non-chaotic models
(Kasner circles) may contain both stable and unstable points, but they
will always oscillate only a finite number of times before arriving in
the set of stable points, after which the behavior is integrable.

\subsubsection{$w > 1$:}
For $w>1$, curvature does not affect the contraction.  The key is the
time-dependence of the $\beta_i$ in (\ref{eq:wg1}), which approach
zero as a positive power of $t$ as $t \to 0$.  Consequently,
the exponential factors $e^{\beta_i}$ in the metric approach
constants. The leading order time-behavior of ${P_a}^b$ is simply that
of a homogeneous and isotropic universe,
\begin{equation}
P \sim \frac{1}{a^2} \sim  |t|^{-\frac{4}{3(1+w)}}.
\label{eq:Pab_divergences}
\end{equation}
This is always less divergent that $t^{-2}$ for $w>1$.
Thus, even in the presence of initial anisotropy and
curvature, the solution for $w > 1$ converges to the isotropic
solution represented by the central point on the Kasner sphere in Fig.~1.

\medskip

We can generalize our cosmic no-hair theorem (described at the end of
section \ref{sec:nocurv}) to include models with spatial
curvature. The Einstein equations for a contracting universe with
anisotropy and inhomogeneous spatial curvature converge to the
Friedmann equation for a homogeneous, flat and isotropic universe if
it contains energy with $w>1$, and that for a homogeneous, flat but
anisotropic universe if $w=1$.  The $w<1$ case becomes highly
inhomogeneous and the no-hair theorem is inapplicable.

\section{Coupling  to p-forms and Chaos}
\label{sec:pforms}

In section \ref{sec:nohair}, we assumed that the evolution of the
universe was dominated by an energy component with fixed equation of
state evolving independently of other matter in the universe. The
component could have been a scalar field or a perfect fluid.  We found
chaotic behavior for $w<1$ in the presence of curvature but
non-chaotic behavior for $w\ge 1$.

In this section, we want to consider how the behavior for $w\ge1$ can
change if the fluid is imperfect or couples to other components.  In
many theories, including Kaluza-Klein, supergravity and superstring
models, the relevant energy consists of a scalar field that is coupled
to $p$-forms. Consequently, we will focus on this important example,
as others have in the past~\cite{Bel73,Dam00,Dam00A,Dam02,Dam02A}.

To determine the effect of the coupling to $p$-forms on chaotic
behavior, our approach is similar to our analysis for spatial
curvature, where we assume an initial state in which the spatial
curvature is negligible and then check that it remains small.  Here we
assume that the $p$-form field strength is initially negligible and
ask how its contribution evolves relative to the energy density with
equation of state $w$. Our action is
\begin{equation}
S=\int
d^4x\sqrt{-g}\bigl(\tfrac{1}{2}R-\tfrac{1}{2}(\partial\phi)^2-V(\phi)
-\tfrac{1}{2(p+1)!}e^{\lambda\phi}{F_{\mu_1\cdots\mu_{p+1}}}^2\bigr),
\label{eq:pformaction}
\end{equation}
where $g$ is the metric, $R$ is the scalar curvature, $V$ is a
potential for the scalar field $\phi$, $p$ is the rank of the $p$-form,
$F$ is the associated field strength tensor and $\lambda$ is the
coupling constant. 
The potential $V(\phi)$ is chosen to give fixed
equation of state $w\ge 1$ in the absence a $p$-form coupling: 
\begin{equation}
V(\phi)=-V_0e^{-\sqrt{3(1+w)}\phi},
\label{eq:potential}
\end{equation}
where $V_0$ is a positive constant. Throughout this paper, we assume
without loss of generality that $\phi\to-\infty$ as $a\to 0$.

For a given equation of state $w$ and $p$-form rank, the behavior of
the system as $t \to 0$ depends on the coupling $\lambda$. We
can extend the terminology introduced earlier to describe the
properties for a given $\lambda$.  We classify the $p$-form coupling
parameter $\lambda$ as \textit{supercritical} if the $p$-form terms
grow relative to the scalar field energy density.  We call these
models supercritical, as opposed to chaotic, because if $w>1$ it is
not known whether chaos occurs or whether the $p$-forms merely play a
non-negligible role in integrable dynamics. In the special case $w=1$,
chaos is known to occur, and we call these models \textit{chaotic}
\cite{Dam00,Dam00A,Dam02,Dam02A}.  Values of $\lambda$ for which the
contracting solution with negligible $p$-forms is stable are called
\textit{non-chaotic} (some authors use \textit{subcritical}). These
two cases are analogous to those introduced in section
\ref{sec:nohair}. If $\lambda$ is on the boundary between
supercritical and non-chaotic, we call $\lambda$
\textit{critical}. The behavior of critical models may be novel, and
will be discussed at the end of this section.

We are assuming that initially the spatial curvature, the anisotropy
and the $p$-form terms are small, and then we check if these
conditions are maintained as the universe contracts. Since we are
considering models where $w\ge 1$, the model is non-chaotic if the
$p$-forms are negligible. The universe may be approximated initially
by the homogeneous isotropic Friedmann-Robertson-Walker form in
(\ref{eq:gen_kasner}) with $\beta_i\approx 0$ and
$\sigma^{(i)}=dx^i$. If $w>1$ and the $p$-form terms are negligible,
$\Omega_\sigma\to 0$ as the crunch approaches. For $w=1$, $\Omega_\sigma$
remains small but finite. If the isotropic case is unstable,
then adding anisotropy cannot restore stability; just as in section
\ref{sec:curv}, the isotropic scale factors are the most stable.

It can be shown that the $p$-form terms involving the spatial
gradients of $F$ grow slower than the leading homogeneous
time-derivative terms, another example of the ultralocal behavior
discussed previously.  Hence, we neglect all spatial derivatives of
the field strength.

The components of $F$ with purely spatial indices, $F_{i_1\cdots
i_{p+1}}$ are called \textit{magnetic} and the components with one
time index, $F^{0i_1\cdots i_p}$, are called
\textit{electric}, in analogy
with the Maxwell action. We will use the labels $E$ 
and $B$ to indicate their respective contributions.
$F$ has a vanishing
exterior derivative $dF=0$. In coordinate notation, 
neglecting the spatial derivatives of $F$, this corresponds to
\begin{equation}
\partial_{[0}F_{i_1\cdots i_{p+1}]}=0,
\end{equation}
where the brackets $[\cdots]$ indicate antisymmetrization.   Thus, the
magnetic components are constant, 
\begin{equation}
F_{i_1\cdots i_{p+1}} = \text{(constant)}
\end{equation}
The equation of motion for $F$ is
\begin{equation}
\nabla_\mu(e^{\lambda\phi}F^{\mu\mu_2\cdots\mu_{p+1}})=
\partial_\mu(e^{\lambda\phi}F^{\mu\mu_2\cdots\mu_{p+1}})+{\Gamma^\mu}_{\mu\sigma}e
^{\lambda\phi}F^{\sigma\mu_2\cdots\mu_{p+1}}=0.
\end{equation}
Only one set of Christoffel symbols appears due to the antisymmetry
of $F$. Since ${\Gamma^\mu}_{\mu 0}=\frac{\partial}{\partial
t}\log\sqrt{-g}$ and ${\Gamma^\mu}_{\mu i}=0$, we can integrate to find,
\begin{equation}
F^{0i_1\cdots i_p}=\frac{e^{-\lambda\phi}}{\sqrt{-g}}
\times \text{(constant)}.
\label{eq:electric}
\end{equation}
The $p$-form part of the stress-energy tensor is
\begin{equation}
T_{\mu\nu}=\frac{e^{\lambda\phi}}{(p+1)!}\bigl((p+1)F_{\mu\mu_2\cdots\mu_{p+1}}{F_
\nu}^{\mu_2\cdots\mu_{p+1}}-\tfrac{1}{2}g_{\mu\nu} F^2\bigr).
\label{eq:p-form_T}
\end{equation}
Decomposing (\ref{eq:p-form_T}) into electric and magnetic components,
and including factors of the metric, we compute the energy density for
the $p$-forms $\rho_p = -{T_{0}}^0$, which is,
\begin{align}
\rho_p &=\frac{e^{\lambda\phi}}{(p+1)!}\bigl(
\tfrac{p+1}{2}F^{0i_1\cdots i_p}F^{0j_1\cdots j_p}g_{i_1j_1}\cdots g_{i_pj_p}
+\tfrac{1}{2}F_{i_1\cdots i_{p+1}}F_{j_1\cdots j_{p+1}}g^{i_1j_1}\cdots 
g^{i_{p+1}j_{p+1}}\bigr)\\
&=\frac{e^{-\lambda\phi}}{a^{2(3-
p)}}\alpha_E^2+\frac{e^{\lambda\phi}}{a^{2(p+1)}}\alpha_B^2,
\label{eq:T00}
\end{align}
where the positive constants $\alpha_E^2$ and $\alpha_B^2$ represent
the magnitude of the electric and magnetic energy, respectively. We
can now define a new set of fractional energy densities,
\begin{subequations}
\begin{align}
\Omega_\phi&=\rho^{-1}\bigl(\dot{\phi}^2/2+V(\phi)\bigr),\\
\Omega_E&=\rho^{-1}\frac{e^{-\lambda\phi}}{a^{2(3-p)}}\alpha_E^2,\\
\Omega_B&=\rho^{-1}\frac{e^{\lambda\phi}}{a^{2(p+1)}}\alpha_B^2,\\
\rho&=\tfrac{1}{2}\dot{\phi}^2+V(\phi)
+\frac{e^{-\lambda\phi}}{a^{2(3-p)}}\alpha_E^2
+\frac{e^{\lambda\phi}}{a^{2(p+1)}}\alpha_B^2.
\label{eq:rho}
\end{align}
\label{eq:EMomegas}
\end{subequations}
where $\Omega_\phi$ is the energy density in the scalar field and
$\Omega_E$ and $\Omega_M$ are the energy densities in electric and
magnetic modes. We are assuming that the anisotropy is negligible, so
$\Omega_\sigma\approx 0$.

The solution for a $\phi$--dominated universe with equation of state $w$ is,
\begin{equation}
\phi=q\ln\lvert t\rvert,\qquad q=\sqrt{\frac{4}{3(1+w)}},
\end{equation}
and $a = |t/t_0|^{2/3(1+w)}$. Substituting in (\ref{eq:T00}), two
terms in $\rho_p$ may be written as
\begin{equation}
\rho_p=\alpha_E^2\lvert t\rvert^{p_E}+\alpha_B^2\lvert t\rvert^{p_B},
\end{equation}
where $p_E$ and $p_B$ are called the electric and magnetic exponents,
respectively.
They are,
\begin{align}
p_E&=-\frac{4(3-p)}{3(1+w)}-\lambda\sqrt{\frac{4}{3(1+w)}},\\
p_B&=-\frac{4(p+1)}{3(1+w)}+\lambda\sqrt{\frac{4}{3(1+w)}}.
\end{align}
 Note that these expressions are invariant under a duality
transformation, which takes $p \to 2-p$, interchanges the electric and
magnetic modes, and takes $\phi \to -\phi$.

In the Friedmann equation, the scalar field energy density scales as
$t^{-2}$. Consequently, $\Omega_\phi\to 1$ and $\Omega_{E,B}\to 0$ as
the universe contracts if both $p_E$ and $p_B$ are both greater than
$-2$. In this case, the $p$-form contribution is negligible and
$\lambda$ is non-chaotic.  Alternatively, if either $p_E$ or $p_B$ is
less than $-2$, the respective $p$-form terms become large and alter
the dynamics.

\begin{figure}
\epsfig{file=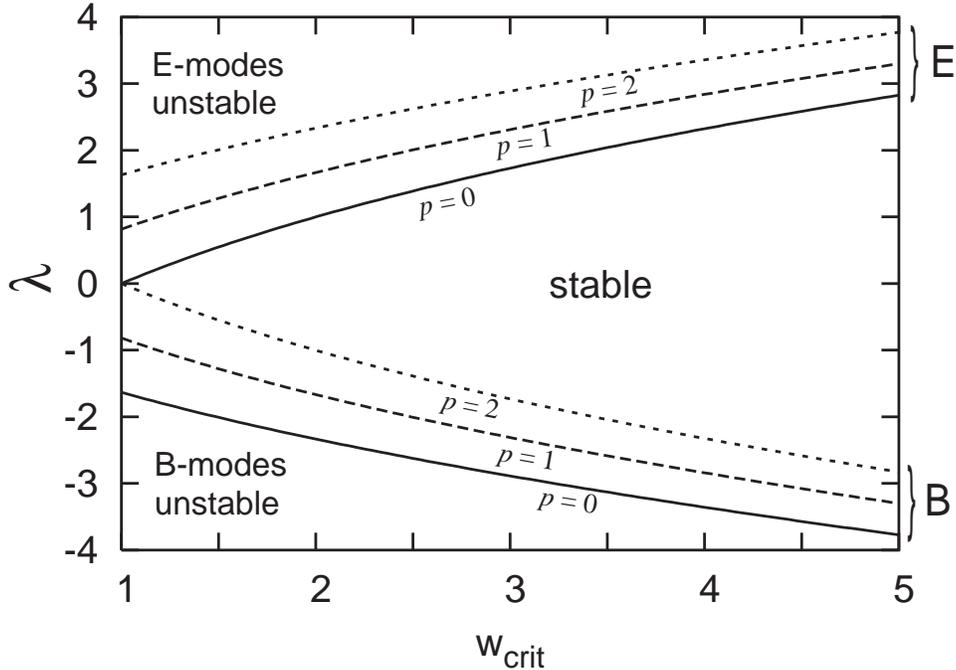,width=5in}
\caption{The four dimensional electric and magnetic couplings
$\lambda$ as a function of the critical equation of state for
$p=0,1,2$. The upper and lower three curves represent the critical
electric and magnetic exponents, respectively. A form with given $p$
and $\lambda$ is stable in a universe with equation of state $w$ if
the point $(w,\lambda)$ lies between the two curves for the given $p$.}
\label{fig:crit}
\end{figure}

For $w=1$, the non-chaotic values of $\lambda$ are
\begin{equation}
\left\{
\begin{aligned}
-\sqrt{8/3}&<\lambda<0 &\qquad p=0\\
-\sqrt{2/3}&<\lambda<\sqrt{2/3}&\qquad p=1\\
0&<\lambda<\sqrt{8/3}&\qquad p=2\\
\end{aligned}
\right.
\end{equation}
Increasing $w$ causes the interval of non-chaotic couplings to grow,
as shown in Fig.~\ref{fig:crit}.  In particular, for any $p$ and
$\lambda$, there exists a critical value $w_{\text{crit}}(\lambda,p)$
such that, for $w>w_{\text{crit}}(\lambda,p)$ the $p$-form terms
remains negligible. For any set of $p$-forms and couplings there
exists a $\bar{w}_{\text{crit}}$, the maximum of $1$ (the critical equation
of state for curvature) and the $w_{\text{crit}}(\lambda,p)$ for each
$p$ and $\lambda$. Then the contraction is non-chaotic if
$w>\bar{w}_{\text{crit}}$.

The behavior can be understood in terms of an \textit{effective}
equation of state for the action (\ref{eq:pformaction}), using the
conservation equation
\begin{equation}
\dot{\rho}=-3\frac{\dot{a}}{a}(1+w_{\text{eff}})\rho,
\end{equation}
where $\rho$ is given by (\ref{eq:rho}). Using (\ref{eq:EMomegas}),
the equation of motion for $\phi$ and the Friedmann equation, we find
\begin{equation}
w_{\text{eff}}=w_\phi\Omega_\phi+\tfrac{3-2p}{3}\Omega_E+\tfrac{2p-1}{3}\Omega_B,
\label{eq:weff}
\end{equation}
where 
\begin{equation}
w_\phi=\frac{\dot{\phi}^2/2-V(\phi)}{\dot{\phi}^2/2+V(\phi)}
\label{eq:wphi}
\end{equation}
is the equation of state for the decoupled scalar field and
$\Omega_\phi+\Omega_E+\Omega_B=1$. The expression (\ref{eq:weff}) is
exact, valid for all values of the $\Omega_i$ assuming the background
is homogeneous, flat and isotropic. For the electric and magnetic
contributions, we can introduce $w_E=\frac{3-2p}{3}$ and
$w_B=\frac{2p-1}{3}$, respectively. The $w_{\text{eff}}$ is just the
$\Omega$-weighted average of $w_\phi$, $w_E$ and $w_B$.

All the $\lambda$ dependence of $w_{\text{eff}}$ is contained in the
time evolution of the $\Omega_i$; $w_\phi$, $w_E$ and $w_B$ do not
depend on $\lambda$. Both $w_E$ and $w_B$ are always less than or
equal to unity, and at least one is strictly less.

If the $p$-form coupling $\lambda$ is non-chaotic, the behavior is
simple. The quantities $\Omega_E$ and $\Omega_B$ rapidly approach zero
as $\Omega_\phi$ approaches one, and the universe is dominated by the
scalar field, with the equation of state $w_\phi$. This is the
non-chaotic case, discussed in section \ref{sec:pforms}.

Alternatively, if the $p$-form coupling is supercritical, $\Omega_E$
and $\Omega_B$ grow. The averaging of the $E$ and
$B$ component ensures $w_{\text{eff}}<w_\phi$. If $w_\phi=1$ then
$w_{\text{eff}}<1$. In this case, the anisotropy grows and chaotic
oscillations occur. It is not known if this happens in the
$w_\phi>1$ case. If in addition, the $p$-form coupling is critical
(so $w_{\text{crit}}=1$), it turns out that the model is equivalent
to an infinite-dimensional hyperbolic Toda system. There are an
infinite number of jumps from one Kasner-like solution to the next,
but the system may be formally integrable \cite{Dam02,Dam02A}.  It is
not clear what the physical ramifications of this behavior are.

\section{Time-varying equation of state and chaos}\label{sec:time_vary}

In a realistic cosmological model, the equation of state will not be
constant, but will depend on the scale factor and approach some
limiting value $w\to\bar{w}$ as $a\to 0$. If $\bar{w}\ne
w_{\text{crit}}$, none of the above analysis changes
substantially. The model is supercritical if $\bar{w}<w_{\text{crit}}$
or non-chaotic if $\bar{w}>w_{\text{crit}}$. The critical case,
$\bar{w}=w_{\text{crit}}$, is more subtle, and the time dependence of
$\bar{w}$ can be significant. In this section, we assume
$w_{\text{crit}}=1$, as this is the most important case, and analyze
what happens when $w_\phi\to 1$ at the crunch. We can expand $w_\phi$
as
\begin{equation}
w_\phi(a)=1+\gamma(a),
\label{eq:gamma}
\end{equation}
where $\gamma$ is a small function of the scale factor such that
$\gamma\to0$ as $a\to0$.

If there is no $p$-form with critical coupling, then using
(\ref{eq:conservation}) and (\ref{eq:beta_w_of_a}), it can be shown
that if $\gamma(a)\log a$ approaches a constant as $a\to 0$, then the
behavior is essentially the same as the $w=1$ case, \textit{i.e.} non-chaotic.
The radius of the Kasner circle in figure \ref{fig:kasner} shrinks, if
$w\to 1^+$, or expands, if $w\to 1^-$.  If $\gamma\to 0$ so
slowly that $\gamma(a)\log a$ diverges as $a\to 0$, then the
anisotropy is eliminated if $\gamma$ approaches zero from above or the
chaos is restored if $\gamma$ approaches from below.

Alternatively, if the model has a $p$-form with critical coupling, the
Kasner contraction will be stable if the $p$-form contribution to the
equations of motion remain subdominant, or, equivalently, if the ratio
of the $p$-form terms to the other terms vanishes in the $a\to 0$
limit. For magnetic modes with critical coupling
$\lambda_{\text{crit}}<0$, we find:
\begin{equation}
\log\frac{\Omega_M}{\Omega_\phi}=\log\frac{e^{\lambda_{\text{crit}}\phi}/a^{2(p+1)}}{\dot{\phi}^2/2+V(\phi)}\sim-
C\int_a^{a_0} \frac{da'}{a'}\gamma(a'),
\label{eq:pformsup}
\end{equation}
where $C$ is a positive constant and by $\sim$ we mean up to terms
finite in the $a\to0$ limit. The behavior is identical if the electric
modes have critical coupling. If $\gamma\to 0$ very slowly, for
example
\begin{equation}
\gamma(a)\sim\lvert\log a\rvert^{-1}
\label{eq:slowepsilon}
\end{equation}
so that the integral diverges as $a\to 0$, then the ratio goes to zero
and $\Omega_M$ becomes negligible in the $a\to 0$ limit. This ensures
that the term is small, and never grows to influence the dynamics.

Let us investigate what conditions on the potential will give us a
$\gamma$ of this form. If we combine the Friedmann equation and
equation of motion for $\phi$, we obtain
\begin{equation}
\frac{d\psi}{d\log a}=3\Bigl(\psi-\frac{V_{,\phi}}{\sqrt{6}V}\Bigr)(\psi-1)(\psi+1),
\label{eq:psieom}
\end{equation}
where $,\phi$ denotes a derivative by $\phi$ and
\begin{equation}
\sqrt{6}\psi=\frac{d\phi}{d\log a}.
\end{equation}
The equation of state (\ref{eq:wphi}) can be expressed in terms of $\psi$,
\begin{equation}
w_\phi=1+\gamma=2\psi^2-1.
\label{eq:wpsi}
\end{equation}

We can obtain $w_\phi\to 1^{+}$ as $a\to 0$ for any
negative potential which is bounded (for large negative values of
$\phi$) by $-Ce^{-\sqrt{6}\phi}$, where $C$ is a positive constant
(see (\ref{eq:potential})). The kinetic energy increases more rapidly
than the potential energy in these cases, and so $w_\phi$ approaches
unity at the crunch. In particular, the potential need not be bounded
below. In general, any potential which can be expressed in the form
\begin{equation}
V(\phi)=2W'(\phi)^2-3W(\phi)^2
\label{eq:posenergy}
\end{equation}
satisfies positive energy \cite{Bou84}. Hertog et al \cite{Her03} have
shown that the potential
\begin{equation}
-V_0 e^{-c\phi},
\label{eq:potential2}
\end{equation}
where $V_0$ and $c$ are positive constants, can be expressed in this
form provided $c<\sqrt{6}$, and so satisfies positive energy. For
$c\ge\sqrt{6}$, solutions exist with total ADM energy that is
unbounded below.

For the potential (\ref{eq:potential2}), $V_{,\phi}/V=c$. In the case
$c<\sqrt{6}$, we find
\begin{equation}
\gamma\propto a^y,
\end{equation}
where $y$ is a positive constant. Consequently, $\gamma\log a\to 0$ as
$a\to0$ and the $p$-form with critical coupling is not
suppressed. However, when $c=\sqrt{6}$, the solution to the equations
of motion show that $\gamma\log a$ approaches a constant, so the
$p$-form can be suppressed when positive energy is violated.

The potential
\begin{equation}
V(\phi)=-V_0e^{-\sqrt{6}\phi}\lvert\phi\rvert^n,
\label{eq:potential3}
\end{equation}
(or more generally, an exponential times any finite order polynomial)
satisfies positive energy (\textit{i.e.}~can be expressed in the form
(\ref{eq:posenergy})) for $n\le-1$. Solving the equation of motion
(\ref{eq:psieom}) for large $\phi$, we find that for $n\le-1$ the
$p$-form with critical coupling is not suppressed. Surprisingly, for
$n>-1$ the ratio (\ref{eq:pformsup}) goes to zero, and the solution is
stable. For the broad class of potentials (\ref{eq:potential2}) and
(\ref{eq:potential3}), the parameters for which they satisfy positive
energy turn out to be exactly those which do not suppress the
$p$-form. It is an open question whether any potential can be
constructed which will suppress the $w_{\text{crit}}=1$ $p$-form and
satisfy positive energy.

\section{Extra Dimensions, Orbifolds and Chaos}
\label{sec:extraD}

In models in which gravity is fundamentally higher dimensional, the
detailed global structure of the extra dimensions can suppress or
enhance chaos in the four dimensional theory.  We consider two simple
compactifications of five dimensional gravity, on $S^1$ and
$S^1/\mathbb{Z}_2$.  In the first, the chaotic nature of pure five
dimensional gravity 
descends to the four dimensional theory.  In the second, the chaotic
behavior is suppressed.  Models of quantum gravity also generally
include additional matter fields in the extra dimensions.  As an
example, we discuss the compactification of heterotic $M$-theory to
four dimensions and find that its behavior during gravitational
contraction is on the borderline between smooth and chaotic.

Consider a five--dimensional, flat universe without matter fields.  We
know from the study of general Kasner universes \cite{Dem85} that it
will exhibit chaotic behavior.  Now compactify one dimension on $S^1$.
We know that the four--dimensional effective theory describes Einstein
gravity coupled to a free scalar field.  The scalar field describes
the volume of the $S^1$ -- it is a simple example of a moduli field.
As all of our preceding arguments regarding gravitational contraction
are local in nature, we expect that the resulting system should be
chaotic as well.
However, a free scalar field has equation of state $w=1$.  According
to our analysis in section \ref{sec:nohair}, one might think that the
behavior should be non-chaotic.  What has happened to the chaos?

The resolution lies in the fact that we have neglected many of the
degrees of freedom of the higher--dimensional theory.  A general 
five dimensional metric $G_{MN}$ can be written,
\begin{equation}
G =
\begin{pmatrix}
h_{\mu\nu}+e^{2q\phi}A_\mu A_\nu & e^{2q\phi}A_\nu \\
e^{2q\phi}A_\mu & e^{2q\phi}
\end{pmatrix}.
\label{eq:kkmetric}
\end{equation}
If we neglect the dependence of the metric on the fifth dimension, 
we may integrate out this dimension and perform a conformal transformation to 
canonically normalize the four dimensional Ricci scalar.  The coefficient
$q$ is then chosen to canonically normalize the scalar field kinetic energy in
the resulting action
\begin{equation}
S=\int d^4x\sqrt{-g}\bigl(\tfrac{1}{2}R-\tfrac{1}{2}(\partial\phi)^2-
\tfrac{1}{4}e^{\sqrt{6}\phi}F^2\bigr),
\label{eq:kkaction}
\end{equation}
which describes a vector field coupled to a free scalar and to
gravity.  The coupling $\lambda = \sqrt{6}$ is outside the stable
range for a 1-form in four dimensions. Therefore, the four dimensional
theory is chaotic, as we would have guessed, but we have to include
the interactions with $p$-forms to see that this is so.

Next, instead of compactifying the fifth dimension on $S^1$, let us compactify
on the orbifold $S^1/\mathbb{Z}_2$.  If the coordinate $x^5$ on $S^1$ runs
from $-\pi$ to $\pi$, this orbifold can be realized as $S^1$ together
with the reflection $x^5 \to -x^5$.  This takes $G_{\mu 5} \to - G_{\mu 5}$, or
equivalently $A_{\mu} \to -A_{\mu}$.  Thus, the Kaluza--Klein 
zero--mode vector field $A_\mu$
is absent in the effective action (\ref{eq:kkaction}).  The absence of this
vector field in the effective action thus implies that the four dimensional 
theory is no longer chaotic.

While orbifolding suppresses some gauge fields and $p$-forms that
would cause chaotic behavior, in some models there are additional $p$-forms 
in the bulk.  These $p$--forms, after dimensional reduction, may 
themselves lead to chaotic behavior.
An illustrative example is heterotic $M$-theory,
which includes a three--form field.

The low--energy four dimensional effective action has been evaluated 
perturbatively by
Lukas et al \cite{Luk97}. To zeroth order in the eleven dimensional
gravitational coupling $\kappa$, it is
\begin{equation}
S^{(0)}=\frac{\pi\rho V}{\kappa^2}\int dx^4\sqrt{-g}\bigl(R-(\partial a)^2-
(\partial c)^2-e^{-\sqrt{8/3}c}(\partial\chi)^2-e^{-
\sqrt{8}a}(\partial\sigma)^2\bigr),
\label{eq:mtheory}
\end{equation}
where we have rescaled the fields in Lukas' action so the kinetic
energies are canonically normalized. The scalar field $c$ is the
radion, which governs the brane separation. The Calabi-Yau
volume modulus $a$ and scalar
field $\sigma$ (which comes from the eleven dimensional 3-form) do not
couple to $c$, and so can be ignored. However, the 3-form modulus
$\chi$ couples to $c$ and the exponent is  critical $\lambda=-\sqrt{8/3}$.
Hence, the theory does not lead to stable Kasner contraction.  
Including the first order ($\kappa^{2/3}$) correction to the action does not
 change the result.  

As this theory is critical, it is quite conceivable that higher order
corrections will lead to a different behavior during cosmological
contraction. There are a number of kinds of corrections to
(\ref{eq:mtheory}) that could push the theory away from criticality
and render it either chaotic or non-chaotic; but, it is not
yet known which behavior occurs.

\section{Conclusions}

The new results presented in this paper build on over three decades of
preceding research on the behavior of cosmological models contracting to a
big crunch.  The classic work focused on cases where the equation of
state of the dominant energy component is $w \le 1$ and $w$ is
constant. The essential
results in this case are:
\begin{itemize}
\item For a perfect fluid with $w<1$, the contraction is smooth and
anisotropic  in the absence of curvature and chaotic
mixmaster if there is non-zero curvature.
\item For a perfect fluid with $w=1$, the contraction is
smooth and \textit{anisotropic}  in the absence of curvature. With
curvature, the contraction is \textit{anisotropic} also, although, depending
on the initial anisotropy, the contraction may undergo a finite
number of jumps from one Kasner-like behavior to another.
\item For a free scalar field coupled to $p$-forms with coupling
$e^{\lambda \phi}$, the contraction is \textit{chaotic mixmaster} if
the coupling $\lambda$ is outside a finite interval of non-chaotic
$\lambda$. The {\it mixmaster case is non-integrable} and the {\it
critical case may be integrable}.
\end{itemize}
In this paper, we have extended  this work to include cases
where $w>1$, a situation that arises naturally in some recent
models with a big crunch/big bang transition, such as the cyclic
and ekpyrotic models.
We have added the following results:
\begin{itemize}
\item For  perfect fluid with $w>1$, the contraction is smooth
and converges to {\it isotropic} at the crunch.
The Einstein equations converge to  ultralocal, homogeneous
and isotropic Friedmann equations.
\item For a scalar field coupled to $p$-forms, there exists
a $w_{\text{crit}}$ such that the contraction is smooth and {\it isotropic}
for $w>w_{\text{crit}}$. 
\item If $w$ is time-varying and approaches one from above
sufficiently slowly the contraction is \textit{smooth} and
\textit{non-chaotic}, even in the presence of a $p$-form with critical
equation of state $w_{\text{crit}}=1$.
\item In models with an extra dimension, compactification
generically
produces a scalar field and $p$-forms.
$\mathbb{Z}_2$ orbifolding
forces  some $p$-forms to zero and, thereby, suppresses their
contributions to chaos.
\end{itemize}
In this paper we have studied how chaotic mixmaster behavior may be
suppressed in models involving a big crunch/big bang transition.  In
particular, the ekpyrotic and cyclic models already include some of
the required ingredients including a scalar field with $w>1$ and $\mathbb{Z}_2$
orbifolding. We did not present a complete non-chaotic
string-motivated model, but the considerations reported here will we
hope be helpful in that regard.

\begin{acknowledgments}
We thank G.~Horowitz and T.~Damour for helpful discussions. This work
was supported in part by NSERC of Canada (JKE), by an NSF Graduate
Research Fellowship (DHW), by US Department of Energy Grant
DE-FG02-91ER40671 and by PPARC (NT). PJS is also Keck Distinguished
Visiting Professor at the Institute for Advanced Study with support
from the Wm.~Keck Foundation and the Monell Foundation.
\end{acknowledgments}


\begin{thebibliography}{abc}
\bibitem{note}
Our conventions are $8\pi G=c=1$; Greek
for spacetime indices ($\mu,\nu,\dots=0,1,\dots,d$); 
Latin purely spatial indices ($i,j,\dots=1,2\dots,d$); and 
$(-+\cdots+)$ sign convention for the metric. For simplicity,
three spatial dimensions
assumed unless otherwise
specified.
\bibitem{noteA} In fact, in a homogeneous universe, $T^0{}_i$ scales
as $a^{-3}$. In this case, a the non-comoving component of a scalar
field or perfect fluid never grows to dominate in a cosmology with
$w>0$. For a universe with a scalar field and negative exponential
potential, such as (\ref{eq:potential}), it is possible to show that
inhomogeneous perturbations to the scaling background may be neglected
near the crunch in a contracting universe.
\bibitem{noteB} Conversely, when $w < 1$, $\alpha$ is negative and
the $\beta_i$ grow rapidly.  Thus even if the energy density $\rho$ is
dominant initially, the anisotropy grows and eventually dominates near
the crunch, a result consistent with our earlier analysis.
\end{thebibliography}
\end{document}